\documentclass{appolb}
\usepackage{graphicx}
\usepackage{float}
\usepackage{textcomp, gensymb}
\usepackage{siunitx}
\usepackage{multicol}
\usepackage{multirow}
\usepackage{microtype}
\usepackage{vwcol} 
\usepackage[T1]{fontenc}
\usepackage[utf8]{inputenc}    
\RequirePackage[top=12mm,bottom=12mm,left=30mm,right=30mm,head=12mm,includeheadfoot]{geometry}

\title{FEASIBILITY STUDIES OF DARK PHOTON SEARCHES WITH THE J-PET DETECTOR 
\thanks{Presented at Symposium on new trends in Nuclear and Medical Physics, 18-20 October 2023}}
\author{Justyna Mędrala-Sowa, Elena Perez del Rio
\address{Faculty of Physics, Astronomy and Applied Computer Science, Jagiellonian University, 30-348 Kraków, Poland\\and\\
Centre for Theranostics, Jagiellonian University, 31-501 Kraków, Poland}
\\[3mm]
{Wojciech Krzemień
\address{High Energy Physics Division, National Centre for Nuclear Research, 05-400 Otwock-Świerk, Poland\\and\\
Centre for Theranostics, Jagiellonian University, 31-501 Kraków, Poland\\[3mm]}}
{on behalf of the J-PET collaboration}
}

\begin{document}
\maketitle

\section{Abstract}
The positronium, a bound state of electron and positron is a unique system to perform highly precise tests, due to no hadronic background and precise Quantum Electrodynamics (QED) predictions. Being a system of lepton and antilepton, its properties are precisely described by QED in the Standard Model (SM). The final events topology can be simulated using Monte Carlo techniques. The J-PET detector is a multi-purpose, large acceptance system that is very well-suitable to the studies of positronium decay due to its excellent angular (1$^o$)  and timing resolutions. 
\\
We present preliminary results on the feasibility of searching for Dark Matter (DM) candidates in the decay o-Ps $\rightarrow$ invisible with the J-PET, which is well suited for the detection of positronium decay products. Toy Monte Carlo simulations have been prepared to incorporate DM decay models to the o-Ps decay expectations in order to assess the detector capabilities to search for such an elusive component of our Universe.

\section{Dark Matter}
\begin{figure}[H]
    \centering
    \includegraphics[width=0.6\textwidth]{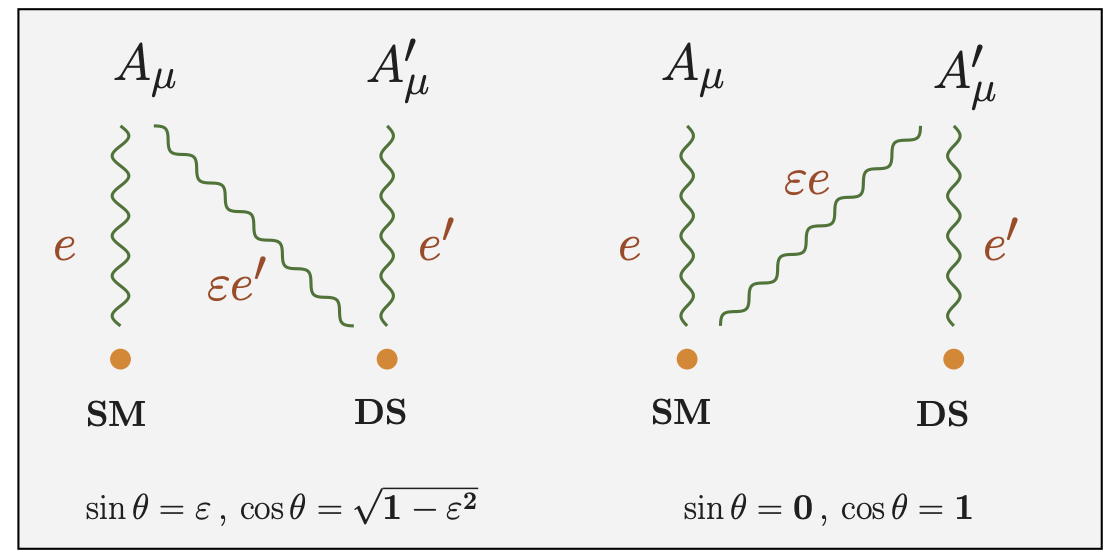}
    \caption{Diagram showing the coupling of an ordinary photon ($A_\mu$) and a dark photon ($A'_\mu$) with SM and dark sector particles. $e$ and $e'$ are the couplings of normal photons and dark photons, respectively \cite{Fabbrichesi_2021}.}
    \label{fig:coupling}
\end{figure}

Dark matter and dark energy play a crucial role in our understanding of cosmology nowadays. The history of discoveries in this area dates back to the 1930s when Zwicky conducted a study of the motion of the galaxy cluster using the virial theorem \cite{ALEXEEV2017141}. 
Nevertheless, during that period, the measurements' methods were not accurate, leading to a lack of credibility for the research. A significant breakthrough came when the Vera Rubin group undertook an exploration of rotation curves, which illustrate the way the speed of stars and gas in a galaxy alters with respect to their distance from the galactic centre \cite{1978ApJ...225L.107R}.
\newline
Dark matter makes up the majority of the matter in the Universe, so it can be assumed that there is a rich so-called 'hidden sector'. Its minimal realisation is one intermediate particle connecting the known SM and hidden particles. The intermediate particle can be a new boson from outside the SM or it can be a new particle within the SM. The proposed gauge boson could be a dark photon or a U boson \cite{S.Bass, Fabbrichesi_2021}. For more than a decade various high-energy and cosmological searches have tried without success setting limits to the coupling to the SM \cite{doi:10.1142/S0217751X19300126}.
\newline
Dark matter particles have the potential to engage with regular matter through kinetic mixing involving both the dark and observable bosons. In theory, this interaction is elucidated by a four-dimensional operator that results from multiplying the field strengths of these two bosons. The existence of this operator implies that one boson can transform into the other as it propagates. Kinetic mixing serves as the bridge between dark and visible particles, essentially acting as the portal through which the detection of dark photons becomes feasible \cite{Fabbrichesi_2021}. Schematically, such an interaction is shown in Fig. \ref{fig:coupling}.

\section{Dark Matter in positronium system}
Positronium is the simplest bound system in which electromagnetic and weak interactions play a role. It is accurately described by QED and that makes it a perfect laboratory for testing QED \cite{S.Bass}.\newline
When considering the ground state, one can speak of two states of the system called: ortho-positronium (o-Ps) and para-positronium (p-Ps). The most important feature that differentiates the states is the mode of annihilation due to the charge parity conservation. P-Ps annihilates into an even number of photons and o-Ps annihilates into an odd number greater than one. Ps cannot annihilate into a single photon, according to the principle of conservation of energy and momentum \cite{Berko_Pendleton_1980}. In practice, the decay is dominated by decay into the smallest possible number of photons.
This is because decays into additional photon pairs are suppressed by powers of the fine structure constant, $\alpha \approx \frac{1}{137}$.
\subsection{U boson}
Exotic decay of the Ps into a U boson and a monochromatic photon is possible according to the model proposed by Pierre Fayet and Marc Mezard described in \cite{FAYET1981226}. If the Ps is in the triplet state only the axial part of the U current contributes as explained in ~\cite{FAYET1981226}, leaving us with U being a pseudovector, with C = +.
\begin{multicols}{2} 
\noindent
It is possible to derive the energy of a monochromatic photon emitted in the o-Ps $\rightarrow \gamma U$ decay as a function of the mass of $U$:
\begin{equation} \label{E_gamma}
    E_\gamma = m_e - \frac{m_U^2}{4m_e}
\end{equation}
\columnbreak
    \begin{figure}[H]
    \centering
    \includegraphics[width=0.3\textwidth]{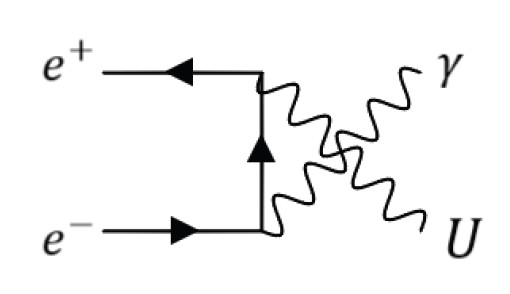}
    \caption{Diagrams of $e^+e^- \rightarrow \gamma U$.}
    \label{fig:my_label}
\end{figure}
\end{multicols}
\noindent
Calculations are performed in the centre-of-mass system of the o-Ps and in the natural units system. 
Estimating the branching ratio of o-Ps $\rightarrow U\gamma$ decay is also possible. The partial lifetime of the ortho-positronium is given by the formula \cite{FAYET1981226}:
\begin{equation}
    \tau (1^3S_1 \rightarrow \gamma U ) \simeq [4/(1-x^4)]\ \si{s};\ x = \frac{m_U}{2m_e}
\end{equation}
Using the usual lifetime for the o-Ps decay $\tau (1^3S_1 \rightarrow \gamma \gamma \gamma ) \simeq 1.4\cdot 10^{-7} s$. The branching ratio for $m_U = 0$ [keV] is then equal to $3.5 \cdot 10^{-8}$. 
\begin{figure}[H]
\centering
\begin{multicols}{2}
    \includegraphics[width=0.9\linewidth]{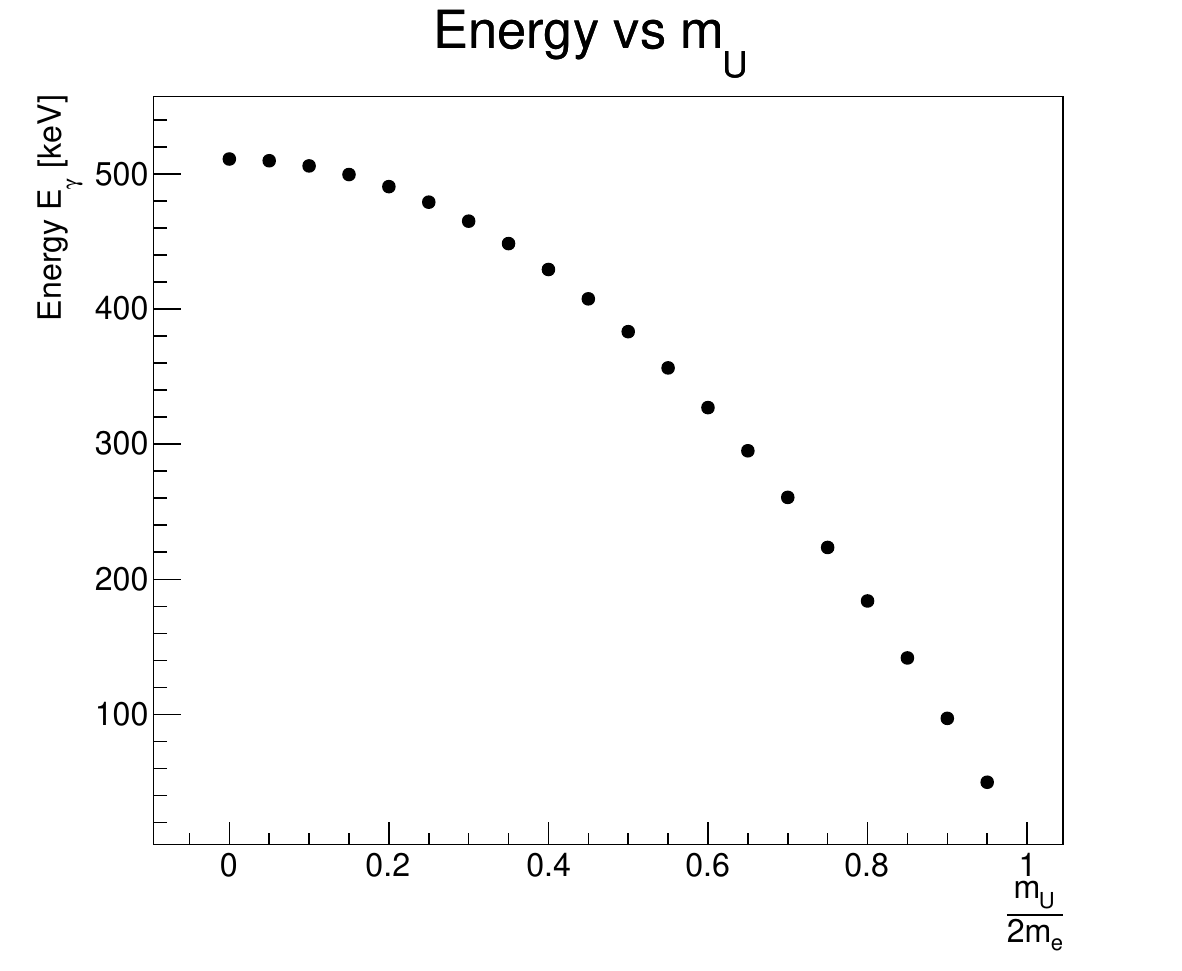}
    \includegraphics[width=0.9\linewidth]{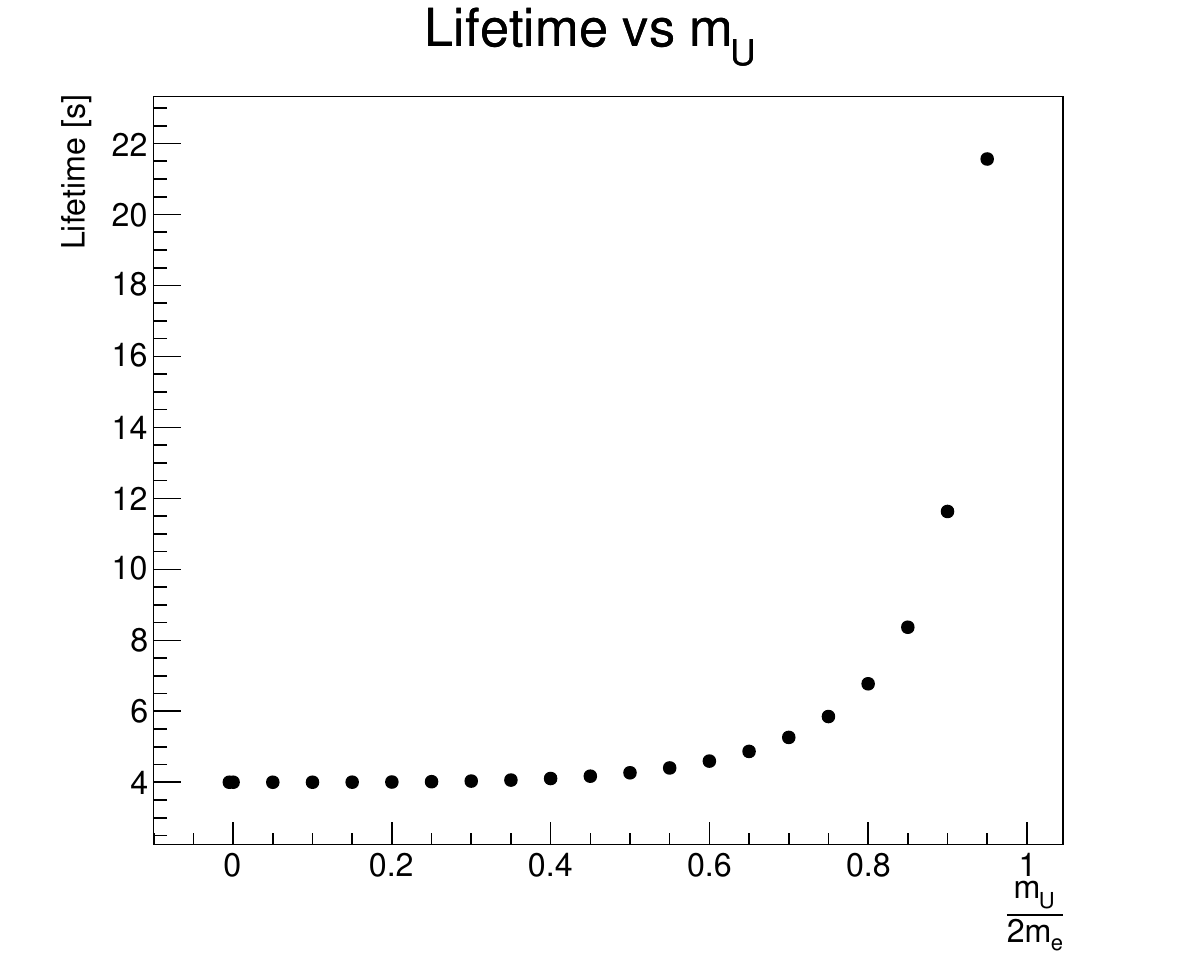} 
    \end{multicols}
    \caption{The energy (left plot) and the lifetime (right plot) of the monochromatic photon emitted in the $o-Ps\rightarrow \gamma U$ decay.}
\end{figure}
\subsection{Background} 
\label{bkg}
The signal from o-Ps $\rightarrow \gamma U$ photon can be mimicked by several other reactions which form a background for this measurement. The largest contribution comes from:
\begin{itemize}
    \item o-Ps $\rightarrow \gamma \gamma \gamma$ for which one photon was recorded, 
    \item p-Ps $\rightarrow \gamma \gamma$ for which one photon was recorded, 
    \item accidental coincidences,
    \item electronic noise.
\end{itemize}
\section{Detector}

J-PET detector was constructed at the Jagiellonian University. It consists of the axially arranged plastic scintillator strips \cite{Moskal_2014}, dedicated programable electronics \cite{Pałka_2017} and a triggerless data acquisition system \cite{Korcyl_2018}. It is a special instrument designed for registering annihilation photons and serves various purposes \cite{detektor2}, including measuring lifetimes, testing discrete symmetries \cite{Moskal_2021}, exploring quantum entanglement \cite{Moskal2018} and medical imaging \cite{jpet1}. Collaborative efforts, such as a search for CPT-violating angular correlation, yield highly precise results \cite{delrio2021physics}.\\
The device is constructed in two stages: the first is the barrel detector, composed of plastic scintillators arranged in three cylindrical layers \cite{detektor1}. A newer, modular design complements this setup, allowing recombination in various configurations with the original detector \cite{in_print}. In the modular J-PET the scintillator strips are is read out by matrices of silicon photomultipliers (SiPM), expected to triple single photon detection efficiency and enhance time resolution by approximately 1.5 times \cite{detektor2}.

The J-PET detector achieved a time resolution of around 380 ps and a position resolution of 4.6 cm (FWHM) \cite{DULSKI2021165452}. It effectively overcomes pile-up events, enhancing the detectability of higher positronium production rates. Pile-up events involve multiple intertwined photons in a single event, and the detector's good resolution mitigates this issue. This capability increases statistical accuracy, providing more precise results in the study of positronium decays \cite{DULSKI2021165452}.
\section{Analysis}

A special chamber is used in the experiment. the chamber is coated with a porous material, which enhances the positronium formation. Among others, the XAD4 polymer \cite{Jasi_ska_2016} was chosen for the experiments. As a result, positronium formation takes place only on the walls of the chamber. The use of XAD4 improves the efficiency of positronium production, and the vacuum at the earlier stages of the positronium's travel ensures that annihilation occurs only with the polymer's electrons.
The J-PET experiment uses as a positron source the isotope $^{22}Na$, which undergoes $\beta ^+$ decay, resulting in excited neon. A photon is then emitted from the excited neon, which is subsequently referred to as the prompt gamma \cite{article}. The mean time of the emission corresponds to 3.7 ps after $\beta ^+$  decay \cite{in_print_Das}. Both processes, the emission of the prompt and the corresponding decay can be considered simultaneous within the detector time resolution.
\newline
 We measure a high-energetic prompt gamma and look for associated one-hit events in the extremely long delayed time window. 
 \begin{equation} \label{eq:N}
     N_{o-Ps \rightarrow \gamma U} (m_U) = \frac{dN}{dt} \cdot BR(m_U) \cdot T \cdot \varepsilon(m_U)
 \end{equation}
where $\frac{dN}{dt}$ - the number of o-Ps produced per second, $BR(m_U)$ - the branching ratio, $T$ - the observation time, $\varepsilon(m_U)$ - the efficiency.\newline
    Four components were taken into account when estimating efficiency:
    \begin{itemize}
        \item geometrical - calculated as a percentage of full solid angle. The detector is cylindrical in shape, but detection only occurs on the sidewalls, not on the bases of the cylinder.
        \item detection - corresponds to the probability of the photon interaction given the thickness of the material.
        \item two components, which aim is to reduce background:
        \begin{itemize}
            \item energy window was considered, where the signal will be the largest and the background the smallest,
            \item time window was considered, where photon from decay $o-Ps \rightarrow U \gamma$ may appear and photon from decay $o-Ps \rightarrow \gamma \gamma \gamma$ is unlikely to be register.
        \end{itemize}
    \end{itemize}

\section{Results}
In the analysis, each component of the formula \ref{eq:N}
was estimated. The following table summarises the values for 4 selected masses of the U boson.
\begin{table}[H]
\caption{The values of selected U boson mass and all parts of the determined efficiency.}
\centering
\begin{tabular}{|cc|cccc|}
\hline
\multicolumn{2}{|c|}{$m_U$ [keV]}                                                                                                                                                            & \multicolumn{1}{c|}{0.0}   & \multicolumn{1}{c|}{255.5} & \multicolumn{1}{c|}{511.0} & 715.4 \\ \hline
\multicolumn{1}{|c|}{\multirow{4}{*}{\rotatebox{90}{efficiency \% \ }}} & geometric                                                                                                          & \multicolumn{4}{c|}{70.71}                                                                       \\ \cline{2-6} 
\multicolumn{1}{|c|}{}                                               & detection                                                                                                              & \multicolumn{1}{c|}{17.47}  & \multicolumn{1}{c|}{17.90}  & \multicolumn{1}{c|}{19.41}  & 22.12  \\ \cline{2-6} 
\multicolumn{1}{|c|}{}                                               & contribution from deposited energy                                                                                    & \multicolumn{1}{c|}{74.52}  & \multicolumn{1}{c|}{72.00}  & \multicolumn{1}{c|}{61.18}  & 0.75   \\ \cline{2-6} 
\multicolumn{1}{|c|}{}                                               & \begin{tabular}[c]{@{}c@{}}contribution from the time window \\ ($t_{shitf}$ = 200 ns, $t_{acc}$ = 50 ms)\end{tabular} & \multicolumn{4}{c|}{1.24}                                                                        \\ \hline
\multicolumn{1}{|c|}{}                                               & total efficiency \%                                                                                                   & \multicolumn{1}{c|}{0.114}  & \multicolumn{1}{c|}{0.113}  & \multicolumn{1}{c|}{0.104}  & 0.001  \\ \hline
\end{tabular}
\end{table}

The reported values allowed the generation of the $N_{o-Ps \rightarrow U\gamma}$ distributions. The uncertainty of each point is equal to $\sqrt{N_{o-Ps \rightarrow U\gamma}}$. Only statistical uncertainties are taken into consideration.
\begin{figure}[H]
\centering
\begin{multicols}{2}
    \includegraphics[width=0.9\linewidth]{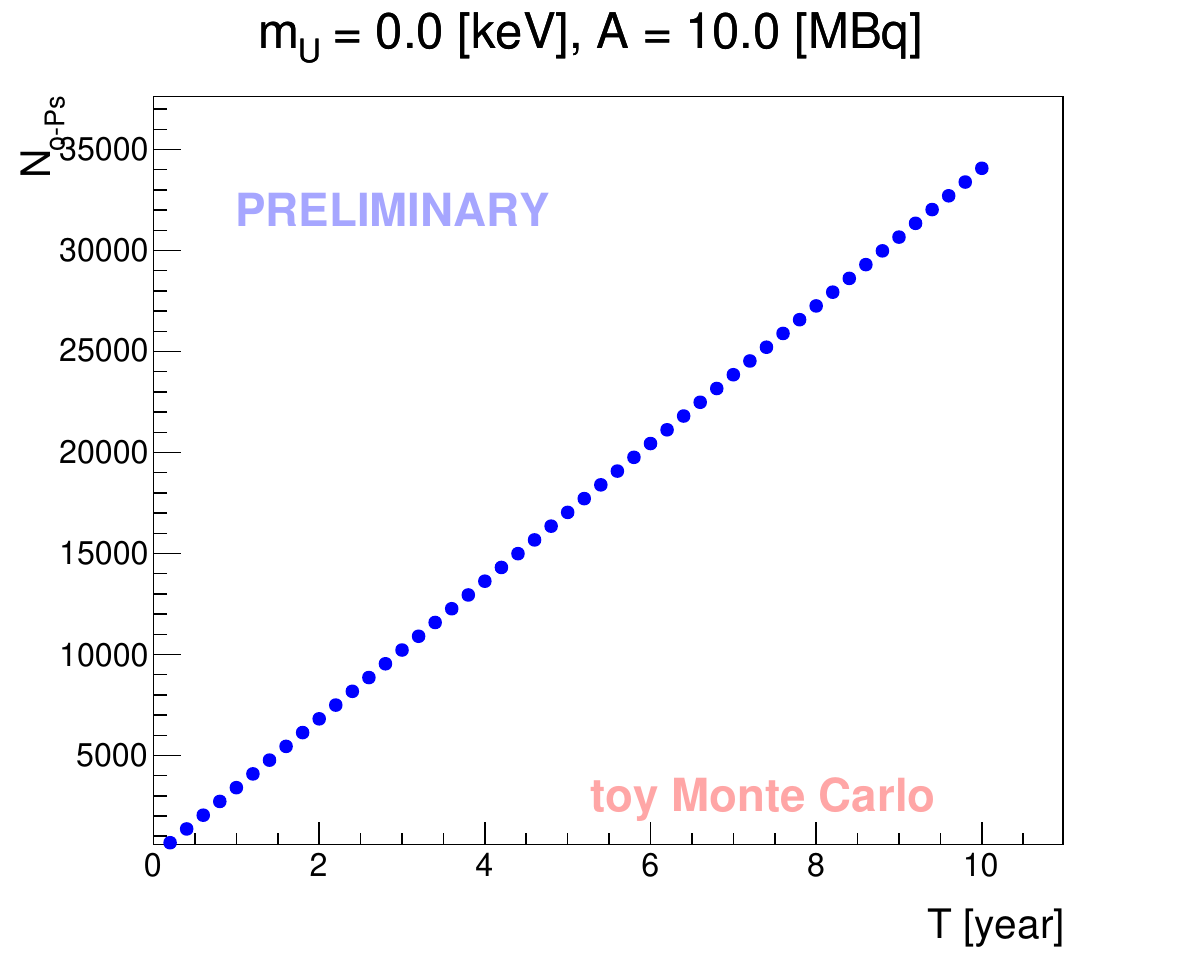}\par
    \includegraphics[width=0.9\linewidth]{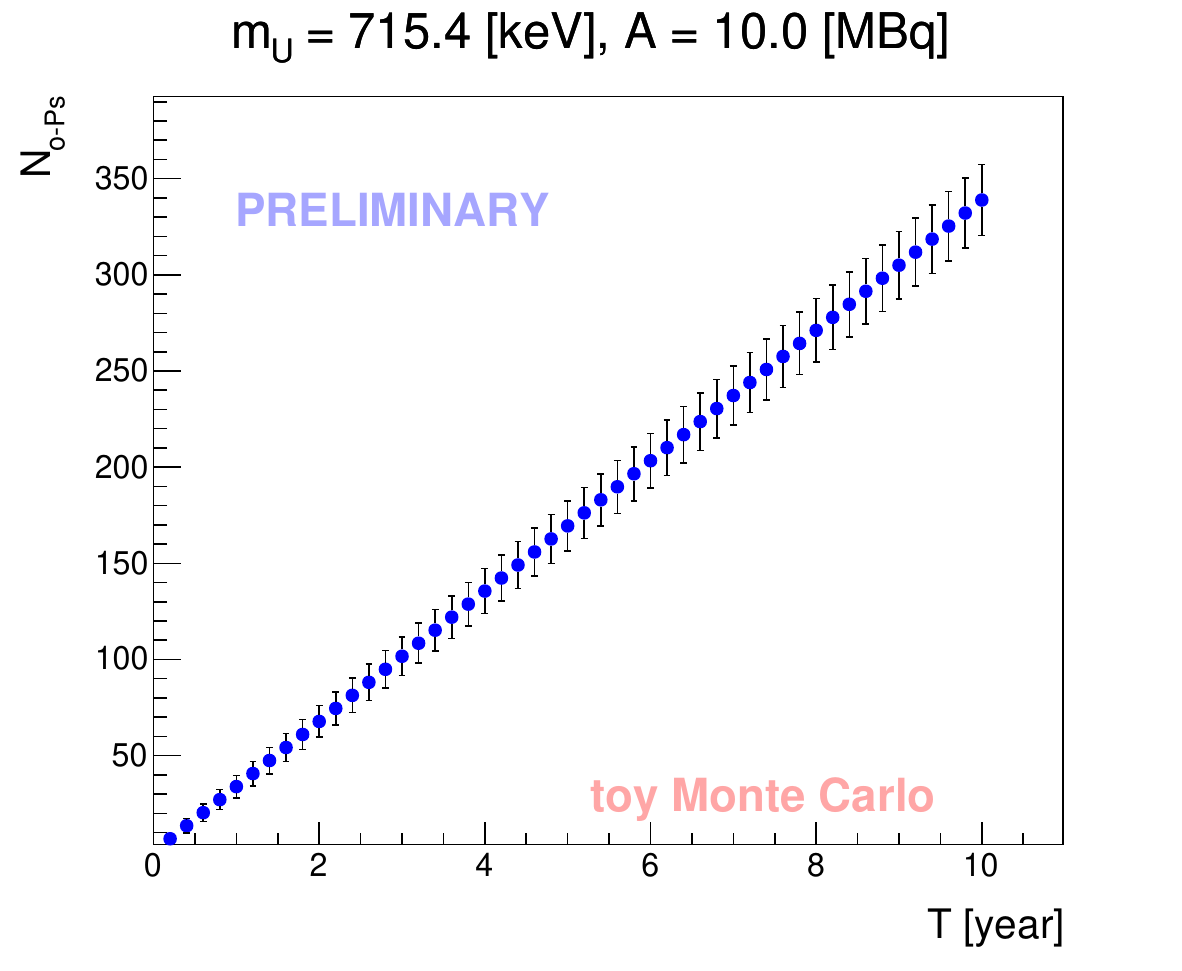}\par 
\end{multicols}
\caption{Number of observed o-Ps $\rightarrow \gamma U$ decays during the observation time for $m_U$ = [0, 715.4] keV and source activity A = 10 MBq. The error bars represent the level of statistical error.}
\end{figure}

\section{Summary and conclusions}
The aim of the research was to assess the possibility of using the J-PET detector for the searches of the Dark Matter candidates in the ortho-positronium decays. The number of ortho-positronia formed in the J-PET's production chamber was analysed. The yield was studied in depth, specifically efficiency.
\newline 
While this work shows that the J-PET is able to perform the search for the U boson, the result evidences that a big improvement can be achieved by further suppressing the background. More refined modeling of the background and detector response, introducing the outcome of this work into the GEANT4 J-PET code \cite{1610988, AGOSTINELLI2003250, ALLISON2016186}, would help to improve the sensitivity.

\section{Acknowledgements}
    We acknowledge support from the National Science Centre of Poland through Grants No. 2019/35/B/ST2/03562, 2020/38/E/ST2/00112, the Ministry of Education and Science through grant no. SPUB/SP/490528/2021, and the SciMat and qLife Priority Research Area budget under the auspices of the program Excellence Initiative-Research University at Jagiellonian University.
\nocite{*} 
\bibliographystyle{unsrt}
\bibliography{bibliography}

\end{document}